\documentstyle[12pt]{article}

\begin{document}
\begin{titlepage}

\title{{\bf String Effective Actions and \\
Cosmological Stability of Scalar Potentials}
\thanks{Work partially supported by CICYT under
contract AEN90--0139.}}

\author{
{\bf Juan Garc\'{\i}a--Bellido}\thanks{Supported by
MEC--FPI Grant. e--mail: bellido@cc.csic.es} \\
Instituto de Estructura de la Materia,  CSIC\\
Serrano, 123 E--28006 Madrid, Spain  \vspace{5mm}\\
{\bf Mariano Quir\'os}\thanks{e--mail: imtma27@cc.csic.es} \\
Theory Division, CERN\\
CH--1211 Geneva 23, Switzerland}

\date{}
\maketitle
\def\baselinestretch{1.15}
\begin{abstract}
The cosmology of the string effective action, including one loop
string threshold corrections, is analyzed for static
compactifications. The stability of the minima of a general
supersymmetry breaking potential is studied in the presence of
radiation. In particular, it is shown that the radiation bath
makes the minima with negative cosmological constant unstable.
\end{abstract}

\vskip6cm
\leftline{CERN--TH.6442/92}
\leftline{IEM--FT--51/92}
\leftline{March 1992}

\vskip-22cm
\rightline{CERN--TH.6442/92}
\rightline{IEM--FT--51/92}
\vskip3in

\end{titlepage}

\newpage
\def\baselinestretch{1.25}

The gravitational sector of closed string theories \cite{GSW} is
a model independent prediction that can be used to check the
effective theory at low energies.
In particular, the dilaton has been recently
recognized to play an essential role in the duality
properties of string theory in the presence of time dependent
backgrounds [2--5]. It also plays a distinghished role in
the cosmological solutions of the gravitational
equations of motion \cite{MUE}, corresponding to a scalar--tensor
theory of gravity \cite{NPB}, and providing a time dependence to
gauge couplings \cite{CQG}. However, at the scale of
supersymmetry breaking, a potential for the dilaton can develop.
In that case, the constancy of gauge couplings is related to
the stability of the minimum of the scalar potential.

In a recent paper, Tseytlin and Vafa \cite{TSV} analyzed the
cosmology of the tree level string effective action including
the dilaton field. They proposed a mechanism, based on the
annihilation of winding modes \cite{BRV}, that allowed the
radiation era to follow from the stringy phase, and studied the
cosmological equations of motion in the absence
of supersymmetry breaking. They showed that the
radiation era was an attractor for certain range of initial
conditions. In this case the dilaton field, and therefore gauge
couplings, remain constant and general relativity is recovered.

In this paper we study the cosmology of the string effective
action including one loop string threshold corrections.
We find that the cosmological equations of motion have
essentially the same form as those of ref.\cite{TSV}. We then
introduce a general supersymmetry breaking potential for the
dilaton and analyze the stability of its minima in
the presence of radiation. In particular we show that
minima with negative cosmological constant are unstable.

We consider a critical ($D=N_c+N+1=10$) heterotic
string compactified to $N=3$ space dimensions in a time
dependent compactification background,
$\ G_{MN} =$diag$ (\bar{g}_{\mu\nu},\ e^{\sigma(t)} \delta_{mn})$.
The string effective action in the Einstein frame can be written
in supergravity form \cite{CRE} as
\begin{equation}
\label{SGR}
{\cal S} = \int d^4x \sqrt{-\bar{g}}\ \left[ \frac{1}{2} \bar{R}
- {\cal G}^{\bar{\imath}}_{\ j}\ \partial_\mu X_i\ \partial^\mu
X^{\bar{\jmath}} - e^{\cal G}({\cal G}_{\bar{\imath}}\
{\cal G}_{i\bar{\jmath}}^{-1} {\cal G}_j - 3)
\right] + {\cal S}_m \hspace{5mm} \ ,
\end{equation}
where $X_i\equiv (S,T)$ are scalar fields in chiral supermultiplets,
$\ {\cal G} = K  + \ln |w|^2$ where $K$ is the K\"{a}hler
potential and $w$ the superpotential, and ${\cal S}_m$ includes
the contribution from the rest of (stringy) matter. The $S$ and
$T$ fields are related to the dilaton and moduli through \cite{WIT}
\begin{equation}
\begin{array}{l}
\label{STF}
Re\ S = e^{3\sigma - 2\phi} \\
Re\ T = e^{\sigma} \hspace{1cm}\ .
\end{array}
\end{equation}

The K\"ahler potential \cite{WIT,MOL}, including one loop string
threshold corrections \cite{DKL}, can be written as \cite{DFK}
\begin{equation}
\label{KP1}
K^{(1)}(S,T) = -\ln \left(S+\bar{S} - 3c \ln (T+\bar{T})\right)
- 3\ln (T+\bar{T}) \hspace{5mm} \ ,
\end{equation}
where $c=\frac{1}{8\pi^2}\sum^3_{i=1}\delta^i_{GS}$
is the Green--Schwarz anomaly \cite{DKL,DFK,CML}.
In terms of the new variable
\begin{equation}
\label{YVA}
Y=S+\bar{S} - 3c \ln (T+\bar{T}) \hspace{5mm} \ ,
\end{equation}
the effective action can be written as
\begin{equation}
\label{YEA}
{\cal S} = \int d^4x \sqrt{-\bar{g}}\ \left[ \frac{1}{2} \bar{R}
- \frac{1}{4} \left(\frac{\partial Y}{Y}\right)^2 - \frac{3}{4}
\left(1 - \frac{c}{Y}\right) (\partial \sigma)^2
- V(Y,\sigma) \right] + {\cal S}_m \hspace{5mm} \ ,
\end{equation}
where we have fixed, for simplicity, {\it Im S} and {\it Im T}
to constant values. (For a cosmological analysis
including a non--trivial {\em Im S}, see ref.\cite{SEY}.)
The action (\ref{YEA}) can be written in the string frame through
a dual invariant \cite{DFK} conformal redefinition of the
four dimensional metric $\bar{g}_{\mu\nu} = Y
g_{\mu\nu}$ \footnote{Note that the action (\ref{EAY}) has the
structure of a Jordan--Brans--Dicke theory \cite{JBD} with
$\omega=-1$ \cite{NPB}.}
\begin{equation}
\label{EAY}
{\cal S} = \int d^4x \sqrt{-g}\ Y \left[ \frac{1}{2} R + \frac{1}{2}
\left(\frac{\partial Y}{Y}\right)^2 - \frac{3}{4} \left(1 -
\frac{c}{Y}\right) (\partial \sigma)^2
- Y V(Y,\sigma) \right] + {\cal S}_m \hspace{5mm} \ .
\end{equation}
We will study the cosmology associated with this theory, with
metric $\ g_{\mu\nu} =$ diag $(-1,\ a^2_i(t) \delta_{ij})$,
$\ i=1,...,N=3\ $, in a similar way as in \cite{TSV}, by defining
the variables
\begin{equation}
\begin{array}{l}
\label{YNV}
\lambda_i(t) = \ln a_i(t) \vspace{1mm}\\
{\displaystyle
\varphi (t) = - \ln Y - \sum^N_{i=1} \ln a_i(t) }
\end{array}
\end{equation}
and writing the effective action as
\begin{equation}
\label{ALE}
{\cal S}=
{\displaystyle
\int dt \sqrt{-g_{oo}}\ e^{-\varphi} \left[
g^{oo} \left( - \sum^N_{i=1} \dot{\lambda}^2_i +
\dot{\varphi}^2 - \frac{3}{2}\left(1 - \frac{c}{Y}\right)
\dot{\sigma}^2 \right) - 2 Y V(Y,\sigma) \right] +{\cal S}_m }.
\end{equation}

{}From now on we will consider static compactification backgrounds
($\sigma=$ constant) and study the cosmological evolution of the
loop corrected string effective action (\ref{ALE}) in the presence
of an arbitrary scalar potential
and a thermal bath of stringy modes with total energy
$E(\lambda)$ and pressure ${\displaystyle P(\lambda)=
-\frac{1}{N}\frac{\partial E(\lambda)}{\partial\lambda} }$.
We take the same scale factor in all ($N$) spatial directions
($\lambda_i\equiv\lambda$).
The action can then be written as
\begin{equation}
\label{LEA}
{\cal S}=
{\displaystyle
\int dt \sqrt{-g_{oo}}\ \left\{e^{-\varphi} \left[
g^{oo} \left( - N \dot{\lambda}^2 +
\dot{\varphi}^2 \right)
-W(Y) \right]-F\right\} },
\end{equation}
where $F$ is the free energy \cite{TSV} and $W(Y)=2YV(Y)$.
The equations of motion derived from the
action ({\ref{LEA}) are
\begin{equation}
\begin{array}{l}
\label{EQM}
\dot{\varphi}^2 - N\dot{\lambda}^2 = e^{\varphi} E + W(Y)
\vspace{1mm}\\
\ddot{\lambda} - \dot{\varphi}\dot{\lambda} =
\frac{1}{2}\ e^{\varphi} P + \frac{1}{2} Y W'(Y) \vspace{2mm}\\
\ddot{\varphi} - N\dot{\lambda}^2 =
\frac{1}{2}\ e^{\varphi} E - \frac{1}{2} Y W'(Y)
\end{array}
\end{equation}
along with the energy conservation equation
\begin{equation}
\label{ECO}
\dot{E} + N\dot{\lambda} P = 0 \hspace{5mm} \ .
\end{equation}

We consider, for large $\lambda$, an energy $E = e^{-\lambda}$
and pressure $P = \frac{1}{N} E$ .\footnote{
In the absence of a potential, see eq.(\ref{EQM}), the solution
$\dot{\varphi} = - N \dot{\lambda} \ $ is an attractor of the
equations of motion \cite{TSV}. This solution exactly
corresponds to a radiation type universe, where $\dot{Y} =
- Y (\dot{\varphi} + N\dot{\lambda}) = 0$, thus leading to constant
gravitational and gauge couplings.}
We can write the equations of motion (\ref{EQM}) as
\begin{equation}
\begin{array}{l}
\label{XYZ}
{\displaystyle
\dot{x} = - \frac{1}{2} x^2 + x y + \frac{1}{2N} y^2
+ \frac{1}{2} W_1(z) } \vspace{2mm}\\
{\displaystyle
\dot{y} = \frac{N}{2} x^2 + \frac{1}{2} y^2
- \frac{1}{2} W_2(z) }
\vspace{2mm}\\
\dot{z} = - z (y + N x)
\end{array}
\end{equation}
where we have defined new variables $y\equiv\dot{\varphi},
\ x\equiv\dot{\lambda}\ $ ($x$ is the Hubble parameter),
$ z\equiv Y$ and the potentials $W_1$ and $W_2$ are given by
\begin{equation}
\begin{array}{l}
\label{W12}
W_1(z) = z W'(z) - \frac{1}{N} W(z) \vspace{1mm}\\
W_2(z) = z W'(z) + W(z) \hspace{1cm} \ .
\end{array}
\end{equation}

Since we are interested in general stability conditions
we have to analyze the non--linear equations of motion
(\ref{XYZ}). The stability of the critical points $x^i_o$ of
first order differential equations
$\dot{x}^i=f^i(x^j)$ is usually determined from
the analysis of the linearized system, $\dot{x}^i= {\cal M}_{ij}
x^j$, where ${\displaystyle {\cal M}_{ij} =
\frac{\partial f^i}{\partial x^j}|_{x^i=x^i_o} }$, and in
particular from the eigenvalues of the matrix ${\cal M}_{ij}$
\footnote{The eigenvalues of this matrix are the slopes of
trajectories in phase space and therefore determine the
stability or instability of those critical points.} \cite{DYN}.
However, in our case the linearized system is not a good
approximation, see eq.(\ref{XYZ}), and we need to develop new
techniques.

Let us first consider the analysis of critical points and stability
of the equations (\ref{XYZ}) in the absence of potential
(for arbitrary $z$). They reduce to
\begin{equation}
\begin{array}{l}
\label{XYN}
{\displaystyle \dot{x} = - \frac{1}{2} x^2 + x y +
\frac{1}{2N} y^2} \vspace{2mm}\\
{\displaystyle
\dot{y} = \frac{N}{2} x^2 + \frac{1}{2} y^2 } \hspace{2cm}\ .
\end{array}
\end{equation}
The only critical point of eq.(\ref{XYN}) is the origin
$(x,y)=(0,0)$. The eigenvalues of the matrix ${\cal M}$
at the critical point are both
$\lambda = 0\ $. Therefore the origin is a star point but its
stability is undetermined \cite{DYN}.

An {\em exact} analytical method for studying this type
of non--linear homogeneous differential equations is based on
the change of variables to polar coordinates $\ x=r\cos\theta\ $
and $\ y=r\sin\theta\ $ so that
${\displaystyle A\equiv \frac{y}{x}= \tan\theta }$.
The stability can then be studied in terms of the parameter
$A$. The equations of motion (\ref{XYN}) can be written as
\begin{equation}
\begin{array}{l}
\label{EPC}
{\displaystyle
\dot{r} = r^2\cos\theta\ \frac{NA(A^2+N+2)+A^2-N}
{2N(1+A^2)} } \vspace{2mm}\\
{\displaystyle
\dot{\theta} = r\cos\theta\ \frac{(N+A)(N-A^2)}{2N(1+A^2)} }
\hspace{1.5cm} \ .
\end{array}
\end{equation}

It is clear from the second equation that there are fixed
directions ($\dot{\theta}=0$) through the origin
for some particular values $A=A_o$. The critical point is
attractive $(\dot{r}<0)$ or repulsive $(\dot{r}>0)$ along these
directions depending on the sign of $y$
\begin{equation}
\begin{array}{l}
\label{ARY}
{\displaystyle
A_o=-N \hspace{8mm} \Longrightarrow \hspace{5mm}
\dot{r}= r y\ \frac{N+1}{2N} } \vspace{2mm}\\
{\displaystyle
A_o=\pm\sqrt{N} \hspace{5mm} \Longrightarrow
\hspace{5mm} \dot{r}=ry } \hspace{1cm} \ .
\end{array}
\end{equation}
Therefore, for $y<0$ the origin is stable while for $y>0$ it is
unstable along the fixed directions $A_o$.
However, we are interested in the stability of the radiation
era, which corresponds to the asymptota $\ y = - Nx$.
The eigenvalue analysis \cite{DYN} only gives information about
the critical points but says nothing about the stability or
instability of the fixed directions ($A=A_o$). We have developed
an analytical method for studying the stability of these
asymptotas, which can be generalized to any non--linear
homogeneous system of first order differential equations.

Let us perturb the angle ($\theta = \theta_o + \epsilon$)
to an infinitesimally close direction $\
A(\epsilon) = \tan(\theta_o + \epsilon) \simeq A_o + \epsilon
(1+A_o^2)$. The angular variation can easily be computed as
\begin{equation}
\label{HPR}
\dot{\theta}(\epsilon)= r\ {\rm sign}(x) \frac{(N+A(\epsilon))
(N-A(\epsilon)^2)}{2N(1+A(\epsilon)^2)^{3/2}} = \ \epsilon\
x\ \frac{N-2NA_o-3A_o^2}{2N} + O(\epsilon^2)\ \ .
\end{equation}
An asymptota ($A=A_o$) will be stable if an infinitesimally
close direction in phase space will approach it, {\em i.e.} if
$\dot{\theta}(\epsilon)$ is negative (positive) for positive
(negative) $\epsilon$. In other words, stable for
$\dot{\theta}\ '(\epsilon=0)<0$ and unstable for
$\dot{\theta}\ '(0)>0$
(where the prime denotes derivative with respect to $\epsilon$).
In our case we have
\begin{equation}
\begin{array}{l}
\label{AXY}
{\displaystyle
A_o=-N \hspace{8mm} \Longrightarrow \hspace{5mm}
\dot{\theta}\ '(0)= -x\ \frac{N-1}{2} } \vspace{2mm}\\
{\displaystyle
A_o=\pm\sqrt{N} \hspace{5mm} \Longrightarrow  \hspace{5mm}
\dot{\theta}\ '(0)=-y\ \left(1\pm\frac{1}{\sqrt{N}}\right) }
\hspace{5mm} \ ,
\end{array}
\end{equation}
and therefore the radiation era ($A_o=-N$) is stable
for $x>0$, in the region between the unstable
asymptotas $\ A_o = \pm \sqrt{N}$, as shown in
the flow diagram of Fig.1, {\em i.e.} the projection of four
dimensional phase space on the plane $(x,y)$
\footnote{In ref.\cite{TSV}
the stability of the radiation era was studied only for initial
conditions in the region $x>0$.}. The
differential equations (\ref{XYN}) determine that the
trajectories in phase space will not escape that region. However,
as we can see from our analysis, initial conditions in other
regions will not converge to the radiation era. In particular,
for $x<0$ the radiation asymptota is {\em unstable}. This result
is crucial for understanding the instability of the minima of a
scalar potential in the presence of radiation.

We now analyze the cosmological evolution described by the
complete equations (\ref{XYZ}). The critical points are \footnote{
Note that for $\ W(0)<0$ there are {\em no} critical points. This
is a general feature of these equations.}
\begin{equation}
\label{CRI}
z=0 \hspace{2mm} \Longrightarrow \hspace{2mm} \left\{
\begin{array}{l}
x=0,\ \ y= +(-)\sqrt{W(0)} \hspace{1cm}
{\rm unstable\ (stable)\ improper \ nodes} \vspace{2mm}\\
{\displaystyle
x=y,\ \ y=\pm\sqrt{\frac{W(0)}{4}} } \hspace{1cm}
{\rm unstable\ saddle\ points}
\end{array} \right.
\end{equation}
and
\begin{equation}
\label{CRP}
y=-Nx \hspace{2mm} \Longrightarrow \hspace{2mm} \left\{
\begin{array}{l}
(N-1) z_o W'(z_o) = 2 W(z_o) \vspace{2mm}\\
{\displaystyle
x_o^2 = \frac{W(z_o)}{N(N-1)} } \vspace{2mm} \\
y_o = - N x_o \hspace{1.5cm}\ .
\end{array} \right.
\end{equation}
The critical points (\ref{CRP}) are the most interesting ones
since they correspond to the radiation era. For $N=3$, the first
condition reduces to
\begin{equation}
\label{CY0}
z_o W'(z_o) - W(z_o) = 2 z_o^2 V'(z_o) = 0
\end{equation}
which gives the extrema of the scalar potential in the
Einstein frame, as one would naturally expect from dynamical
arguments. However, the existence of a critical point at the
radiation asymptota crucially depends on the sign of $W(z_o)$.
Only for positive or zero cosmological constant ($\Lambda \equiv V(z_o)$)
we have a critical point. The solutions of the (cubic)
${\cal M}_{ij}$--eigenvalue equation, for $N=3$,
\begin{equation}
\label{CL3}
\lambda^3 + 7\sqrt{\frac{W(z_o)}{6}} \lambda^2 +
(2 W(z_o) + z_o^2 W''(z_o)) \lambda + 4 \sqrt{\frac{W(z_o)}
{6}} z_o^2 W''(z_o) = 0  \hspace{5mm} \ ,
\end{equation}
determine the stability of the critical points (\ref{CRP}).
A stable center or spiral point corresponds to complex
eigenvalues, while an unstable saddle point corresponds to real
eigenvalues of different sign \cite{DYN}. Therefore, the
condition of stability of these critical points (which are maxima
or minima of the scalar potential $V(z)$) is the existence of
complex solutions to the cubic equation (\ref{CL3}), which gives the
condition on the coefficients, and therefore on the potential,
for $N=3$,
\begin{equation}
\label{CLW}
8 z_o^2 W''(z_o) - 3 W(z_o) > 0 \hspace{5mm} \ .
\end{equation}

This condition can be expressed in terms of the scalar potential
$\ V(z)$ as $\ 8 z_o^2 V''(z_o) > 3 V(z_o)$ and interpreted as
the well known stability condition of extrema of potentials
in a curved background, $\ 4\bar{m}^2 > 3\Lambda\ $
\cite{FRE}, where $\ \bar{m}^2 = 2 z_o^2 V''(z_o)\ $ is the mass of
the scalar field in the Einstein frame.
This result, obtained from the stability analysis of the
gravitational equations of motion at the minimum of the
potential in the Jordan frame, is in agreement
with the result obtained in ref.\cite{FRE}
from the positivity of the total
energy functional of the gravitational system.

So far we have analyzed the stability of the critical points
(\ref{CRP}). Let us now analyze the stability of the radiation
asymptota in the presence of an arbitrary scalar potential.
Suppose that we are driven to the minimum of the potential $\
W(z_o)$. The differential equations (\ref{XYZ}) are homogeneous
and a similar analysis as that of eq.(\ref{HPR}) can be
performed. It gives in this case, for $A_o=-N$,
\begin{equation}
\label{HPO}
\dot{\theta}\ '(0)= -\ \frac{N-1}{2x} \left(x^2 +
\frac{N+1}{(N-1)^2}W(z_o)\right)  \ \ .
\end{equation}
The radiation asymptota is stable (unstable) for large positive
(negative) $ x$, as in the absence of potential (\ref{AXY}).
The equations of motion (\ref{XYZ}) along the radiation
asymptota reduce for $N=3$ to
\begin{equation}
\label{EAN}
\dot{x} = - 2 x^2 + \frac{1}{3} W(z_o) \hspace{5mm} \ .
\end{equation}
We have plotted this function in Fig.2 for different values of
$W(z_o)$. It is a typical example of a structurally unstable
differential equation \cite{DYN}. For $W(z_o)=0$ there is just
one critical point, and the universe will approach it after an
infinite time. For $W(z_o) > 0$ there are two critical points, at
$x_o^{\pm}=\pm\sqrt{\frac{W(z_o)}{6}}$. The positive root is attractive
and the other is repulsive. Assuming initial conditions $x(0)>0$
(as in the zero cosmological constant case) the universe will
approach the attractor at $x=x_o^+$, which corresponds to an
inflationary universe. For $W(z_o) < 0$ there is {\em no}
critical point (as we know from the previous analysis)
and the Hubble parameter decreases until it changes sign
and the universe starts contracting.

For $x(0)>0$, eq.(\ref{EAN}) can be integrated to give the
Hubble parameter and scale factor
\begin{equation}
\begin{array}{l}
\label{SLP}
{\displaystyle
x^{(+)}(t) = \sqrt{\frac{W_o}{6}} \coth 2\sqrt{\frac{W_o}{6}} t }\\
{\displaystyle
a^{(+)}(t) = \left(\frac{1}{2}\sqrt{\frac{W_o}{6}} \sinh 2\sqrt{
\frac{W_o}{6}} t\right)^{1/2} }
\end{array}
\end{equation}
\begin{equation}
\begin{array}{l}
\label{SLZ}
{\displaystyle
x^{(0)}(t) = \frac{1}{2 t} } \vspace{3mm}\\
a^{(0)}(t) = t^{1/2} \vspace{2mm}
\end{array}
\end{equation}
\begin{equation}
\begin{array}{l}
\label{SLN}
{\displaystyle
x^{(-)}(t) = \sqrt{\frac{|W_o|}{6}} \cot 2\sqrt{\frac{|W_o|}{6}} t }\\
{\displaystyle
a^{(-)}(t) = \left(\frac{1}{2}\sqrt{\frac{|W_o|}{6}} \sin 2\sqrt{
\frac{|W_o|}{6}} t\right)^{1/2} }
\end{array}
\end{equation}
that correspond to $W_o\equiv W(z_o)$ positive, zero and negative
respectively. We have plotted these functions in Fig.3. For
$W_o<0$, the universe will start contracting
at a time $\Delta t$ after the beginning of the
radiation era given by
\begin{equation}
\label{DET}
\Delta t = \frac{\pi}{4}\sqrt{\frac{6}{|W_o|}} \hspace{5mm} \ .
\end{equation}
(For a cosmological constant  $\Lambda \sim 10^{-120}$ in Planck
units, we obtain the age of the universe, as expected \cite{COS}).
After the universe has started contracting in the presence of a
radiation bath, the stability condition (\ref{HPO}) determines
that the asymptota $\ A=-N$ is unstable. From the third equation
in (\ref{XYZ}) we deduce that the value of the scalar field $\ Y$
increases (decreases) for $\ y+Nx < 0$ ($> 0$), and escapes from
the minimum of the potential. It is the radiation bath
in an anti--de Sitter background which acts as a source of energy
for the scalar field to pass the potential barrier.

We now apply the previous analysis to a class of supersymmetry
breaking potentials in string theories, namely those provided by
gaugino condensation \cite{GAU}.
The existence of a local minimum in the supersymmetry breaking
potential requires at least two gaugino condensates, and the
scalar potential can be computed from the superpotential
\begin{equation}
\label{SUP}
w(S) = d_1 e^{-\alpha_1 S} + d_2 e^{-\alpha_2 S} \hspace{5mm} \ ,
\end{equation}
where $d_i$ and $\alpha_i$ depend on the gauge group of the hidden
sector. We will not worry about the details but take a generic case
which may be good phenomenologically \cite{BEA}. The scalar
potential will have a local minimum only for non--trivial
imaginary part of the field $S$. In this case the scalar
potential can be written as
\begin{equation}
\begin{array}{lr}
\label{SPO}
Y V(Y) =&d_1^2 e^{-\alpha_1 Y} ((\alpha_1 Y+1)^2-3)
+ d_2^2 e^{-\alpha_2 Y} ((\alpha_2 Y+1)^2-3)  \vspace{2mm}\\
&- 2d_1d_2 e^{-(\alpha_1+\alpha_2) Y/2}
(\alpha_1\alpha_2 Y^2+(\alpha_1+\alpha_2)Y-2) \ .
\end{array}
\end{equation}
where $Y$ is the string loop corrected variable (\ref{YVA}).
These potentials have a {\em negative} cosmological
constant at its local minimum, which can be used to fix the
dilaton and therefore the gauge couplings.
We have plotted $V(Y)$ and $W(Y)$ in Fig.4 for
particular values
of $d_i$ and $\alpha_i$.

We have solved the non--linear equations of motion (\ref{XYZ})
with {\em Mathematica}$^\copyright$ \cite{MAT} and
shown the phase space projection on the $(x,y)$ plane in Fig.5 .
We assumed that the initial conditions are that of radiation at
the minimum of the potential ($y=-3x$, $Y=Y_o$). This is a
reasonable asumption since the scalar potential will begin to be
important at a (condensation) scale much below the Planck scale
and therefore the universe had enough time to converge to the radiation
asymptota. On the other hand, the minimum of the potential is
stable during the expansion of the universe and therefore
the scalar field had time to settle at its minimum.
As predicted from the general analysis (\ref{HPO}),
during the expansion phase the radiation is
stable, but once the universe starts contracting ($x<0$) it
becomes unstable and the solutions diverge from the radiation
asymptota, thus moving the scalar field
away from the minimum of the potential, as shown in Fig.6.
Depending on the initial conditions (above or below the asymptota)
the universe will expand or contract in finite time, while the
scalar field moves towards $\ Y=0$ or $\ Y=\infty$ respectively.
This is shown in Fig.7. For a realistic supersymmetry
breaking potential based on gaugino condensation, the minimum
has a negative cosmological constant of order $\Lambda \sim\
m_{3/2}^2 \sim 10^{-32}$ in Plank units and therefore the
universe would have collapsed by $\Delta t \sim 10^{-27}$ s, much
before primordial nucleosynthesis.

In conclusion, we have studied the cosmology of the string
effective action including one loop string threshold corrections
and analyzed the stability of the minima of a
supersymmetry breaking scalar potential
in the presence of radiation. For positive or zero cosmological
constant the minimum is stable and the universe enters an inflationary
phase or approaches the radiation era, respectively. For negative
cosmological constant the minimum becomes unstable when the universe
starts contracting and the scalar field moves away from the minimum of
the potential.

\section*{Acknowledgements}
We would like to thank A. Valdes for helpful comments on
the stability conditions, and B. De Carlos for supplying us
with some phenomenological values of $\alpha_i$ and $d_i$
from ref.\cite{BEA} prior to publication. We are grateful to F.
Quevedo for discussions on the duality properties of string theory.

\newpage
\section*{Figure Captions}
\begin{description}

\item[Fig.1]
Phase space diagram (solid lines) of equations (\ref{XYN}), in
the projected plane $(x, y)$. The stability of the asymptotas,
$A=\pm\sqrt{3}$ (dashed lines) and $A=-3$ (dashed--dotted line),
in different regions follows the analysis of eq.(\ref{AXY}).

\item[Fig.2]
Phase space $(\dot{x},x)$ of eq.(\ref{EAN})
for $ W_o >0$ (solid curve), $W_o=0$ (dashed curve)
and $W_0<0$ (dashed--dotted curve).
For $\ W_o < 0$ the Hubble parameter $x$ will never
stop decreasing.

\item[Fig.3]
The solutions $\ x(t)$ and $\ a(t)$ (Hubble parameter
and scale factor) of eqs.(\ref{SLP}--\ref{SLN})
for $ W_o >0$ (solid curves), $W_o=0$ (dashed curves)
and $W_0<0$ (dashed--dotted curves). For $\ W_o < 0$ the
scale factor starts contracting after a time
interval $\ \Delta t$ given by eq.(\ref{DET}).

\item[Fig.4]
The scalar potentials $\ V(Y)$ (solid curve) and $\ W(Y)$
(dashed curve), in the Einstein and
Jordan frame respectively, for the hidden gauge group
SU$(4)_1 \times $SU$(5)_2$ as determined by $c=0$,
$\ \alpha_1=2\pi^2,\ \alpha_2=16\pi^2/5$ and $d_1=1/8\pi^2e,\
d_2=5/32\pi^2e\ $ \cite{BEA}. The local minimum corresponds to a
negative cosmological constant. $Y$ is given in units of
$\ \alpha_1$ while $W(Y)$ and $V(Y)$ are in units of $\ d_1^2$.

\item[Fig.5]
Phase space of equations (\ref{XYZ}) in the presence of the
scalar potential $\ W(z)$ of Fig.4 in the projected
plane $(x, y)$, for
initial conditions very close to the radiation asymptota.
As predicted by our analysis, the radiation era is stable in an
expanding universe but {\em unstable} in the contracting phase
($x<0$) which can only be reached for {\em negative} cosmological
constant.

\item[Fig.6]
Phase space projection on the $(z,x)$ plane, for initial
conditions $x=1\pm0.1$, $y=-3$ and $z=z_o$ (very
close to the radiation asymptota) in the presence of a
negative cosmological constant. The minimum of the potential is
unstable since the value of the scalar field decreases to zero
(solid curve) or increases to infinity (dashed curve)
while the Hubble parameter goes to $\ \pm \infty\ $ for
$\ y(0)+3x(0)>0$ or $<0$ respectively.

\item[Fig.7]
Evolution of the scale factor $\ a(t)$ corresponding to the initial
conditions of Fig.6. The universe will expand (solid curve) to
infinity or collapse (dashed curve) in {\em finite} time after
entering the contracting phase.

\end{description}

\end{document}